# An Information Theoretical Analysis of Kinase Activated Phosphorylation Dephosphorylation Cycle

Hong Qian[1] and Sumit Roy[2]

[1] Department of Applied Mathematics
University of Washington, Seattle, WA 98195-2420
qian@amath.washington.edu

[2] Department of Electrical Engineering
University of Washington, Seattle, WA 98195-2500
roy@ee.washington.edu

February 8, 2010

**Abstract**

Signal transduction, the information processing mechanism in biological cells, is carried out by a network of biochemical reactions. The dynamics of driven biochemical reactions can be studied in terms of nonequilibrium statistical physics. Such systems may also be studied in terms of Shannon's information theory. We combine these two perspectives in this study of the basic units (modules) of cellular signaling: the phosphorylation dephosphorylation cycle (PdPC) and the guanosine triphosphatase (GTPase). We show that the channel capacity is zero if and only if the free energy expenditure of biochemical system is zero. In fact, a positive correlation between the channel capacity and free energy expenditure is observed. In terms of the information theory, a linear signaling cascade consisting of multiple steps of PdPC can function as a distributed "multistage code". With increasing number of steps in the cascade, the system trades channel capacity with the code complexity. Our analysis shows that while a static code can be molecular structural based; a biochemical communication channel has to have energy expenditure.

# 1 Introduction

Cellular biochemical signal transductions are communication processes on a molecular level through chemistry. While the medium for the information processing is very different from electronic and optical, the fundamental principles in the theory for communications apply. Currently there is a growing fascination with Shannon's information theory in cellular biochemistry [4, 1, 14, 18].

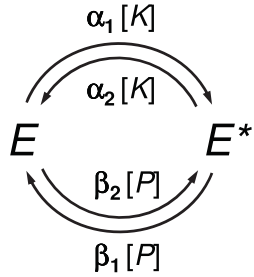

Figure 1: A simple PdPC with non-saturated kinase and phosphatase. $[K]$ and $[P]$ are the concentrations of the kinase and phosphatase. Note that an enzyme has to catalyze both the forward and backward reactions. Dephosphorylation, $E^* \to E + Pi$, is *not* the reverse reaction of the phosphorylation reaction, $E + ATP \to E^* + ADP$. Identical kinetics arises in GTPase signaling, where $E$, $E^*$, $K$, and $P$ correspond to GDP·GTPase, GTP·GTPase, GEF and GAP, respectively (see main text).

Biological information, either originating from DNA or from extra-cellular signals, are encoded in the activities of signaling proteins and delivered as the biochemical activity propagates through reaction pathways. Since all these processes are carried out by molecules, chemical thermodynamic analysis has been developed to quantify the energetic aspect of signal transduction processes. We have recently shown that free energy expenditure is an important, but often overlooked, aspect of cellular signal transduction [20, 21, 22].

To quantify the functional aspect of signal transduction as encoding of biochemical states and transmission of biochemical information over channels, in this paper we introduce a formal Shannon information theoretic analysis [13, 26, 5] into the most basic unit of cellular signal transduction: the phosphorylation-dephosphorylation cycle (PdPC) and the GTPase. These two signaling modules are kinetically isomorphic, with GDP and GTP bound GTPases correspond to the dephosphorylated and phosphorylated states of a protein, and guanine-nucleotide exchange factor (GEF) and GTPase accelerating protein (GAP) correspond to the kinase and the phosphatase [22, 2]. The analysis developed below for the PdPC can be equally applied to the GTPase.

# 2 PdPC Switch as an Informational Transfer Channel

## 2.1 Phosphorylation dephosphorylation cycle (PdPC)

We consider a signaling molecule, a protein, which can be chemically modified via phosphorylation and dephosphorylation, catalyzed by a protein kinase $K$ and protein phosphatase $P$ respectively. Let $E$ and $E^*$ be the dephosphorylated and phosphorylated states of the protein. Assuming that neither the kinase nor the phosphatase are saturated. Hence we have a simple kinetic scheme shown in Fig. 1 [2, 22].

The PdPC shown in Fig. 1 has been extensively studied, from kinetic and thermodynamic perspective, in [20, 22, 23, 8, 2]. When the kinase and the phosphatase are operating in the non-saturated linear regime, the $\alpha$'s and $\beta$'s are the ratio of corresponding $k_{cat}$ to $K_M$. Moreover, the $\alpha_1 = \alpha_1^o[ATP]$, $\alpha_2 = \alpha_2^o[ADP]$, and $\beta_2 = \beta_2^o[Pi]$. Therefore,

$$\gamma = \frac{\alpha_1 \beta_1}{\alpha_2 \beta_2} = \frac{K_{ATP}[ATP]}{[ADP][Pi]} = e^{\Delta G/RT}, \tag{1}$$

where $K_{ATP}$ is the equilibrium constant of the ATP hydrolysis reaction; $\Delta G$ is the ATP hydrolysis free energy. Note that the hydrolysis free energy is in the sustained high concentration of ATP and low concentrations of ADP and Pi, away from their chemical equilibrium. The useful hydrolysis free energy is *not* in the phosphate bond of the ATP molecule.

The fraction of the protein in the phosphorylated state $E^*$ is [21, 22, 2]

$$\begin{aligned} f &= \frac{[E^*]}{[E] + [E^*]} \\ &= \frac{\alpha_1[K] + \beta_2[P]}{\alpha_1[K] + \alpha_2[K] + \beta_1[P] + \beta_2[P]} \\ &= \frac{\theta + \mu}{1 + \theta + \mu + \theta/(\gamma\mu)}, \end{aligned} \tag{2}$$

where

$$\theta = \frac{\alpha_1[K]}{\beta_1[P]}, \quad \mu = \frac{\beta_2}{\beta_1}, \quad \text{and} \quad \gamma = \frac{\alpha_1 \beta_1}{\alpha_2 \beta_2}. \tag{3}$$

The parameter $\theta$ characterizes the level of "upstream signal", the parameter $\mu$ characterizes the level of background in the absence of the kinase, and the parameter $\gamma$ characterizes the amount of energy available from ATP hydrolysis. The $\Delta G$ in Eq. (1) is known as the phosphorylation potential, which is different from the standard state free energy $\Delta G^o = RT \ln K_{ATP}$. A normal cell has $\Delta G$ on the order of 13 kcal/mol, which corresponds to $\gamma = 10^{10}$.

From a thermodynamic perspective of the biochemical reactions, there is always a certain amount of phosphorlation even without the kinase. And likewise, there is always less than 100% of phosphorylation even with the presence of the kinase. This is the physical origin of the 'channel noise' from information theoretical perspective.

## 2.2 Amplitude of the switch and information transfer matrix

So far, we have presented the PdPC as a biochemical kinetic system. We now consider it as a logical system for information transfer, i.e., signaling transduction.

The PdPC is widely considered as a 'switch' in cellular biology: An increase in the kinase concentration (or activity) will lead to the phosphorylation of $E \to E^*$. We now define the *amplitude of the switch* ($AoS$) [21]:

$$AoS = f_{\theta=\infty} - f_{\theta=0} = \frac{\gamma\mu}{1+\gamma\mu} - \frac{\mu}{1+\mu}. \tag{4}$$

In fact, characteristics of PdPC activation by the kinase can be represented by a matrix

$$\mathbf{K} = \begin{pmatrix} \frac{1}{1+\mu} & \frac{\mu}{1+\mu} \\ \frac{1}{\gamma\mu+1} & \frac{\gamma\mu}{\gamma\mu+1} \end{pmatrix} \tag{5}$$

where $\frac{\mu}{1+\mu}$ is the probability of the signaling protein being activated in the complete absence of its kinase; and $\frac{\gamma\mu}{1+\gamma\mu}$ is the probability in the presence of sufficiently amount of kinase.

It has been shown that the amplitude of the switch is independent of whether the kinase and phosphatase are operating under the linear or saturated regimes [22, 2]. Therefore, the results obtained in our present work apply equally well to PdPC with saturated enzyme kinetics.

## 2.3 Information theory

The matrix $\mathbf{K}$ in Eq. (5) is analogous to the transition probability matrix relating the input ($X$) and output ($Y$) random variables, of a discrete, memoryless channel. This forms the bedrock of the theory of information transmission over a noisy channel, developed by Shannon [13, 26, 5]. We provide a very brief discussion of this theory for the case of binary inputs and outputs as is appropriate for this scenario. Without loss of generality, we label the input/output states as $0, 1$ respectively; hence the elements of the matrix $\mathbf{K}$ in Eq. (5) can be interpreted as the conditional probabilities $q_{j,i} = P\{Y = y_j | X = x_i\}$ where $y_j, x_i = \{0, 1\}$.

Shannon introduced the notion of *entropy* to capture the average uncertainty in a discrete random variable $X$, defined by $H(X) = -\sum_i p_i \ln p_i$ where the random variable $X = x_i$ with probability $p_i$. For a binary $(0, 1)$ random variable $X$, this is

$$H(X) = -p_0 \ln p_0 - (1-p_0) \ln(1-p_0) \tag{6}$$

where $p_0 = \Pr\{X = 0\}$. In the theory of (noiseless) source coding [5], the entropy $H(X)$ equals the number of bits/symbol required to represent and transmit a sequence of symbols chosen randomly and independently from the source $X$; clearly this is at maximum, equal to $\log_2 |X|$ where $|X|$ is the cardinality of $X$. However, as we have seen, the PdPC is noisy and we apply the theory of information transmission over a noisy channel [5].

Shannon's information theory provides a limit - called the channel *capacity* - on the information carrying capability of any such noisy channel, that represents the *maximum rate* (in bits/symbol or channel use) at which the source symbols may be communicated with arbitrarily low rates of error over such a channel. This is best understood in terms of the concept of *mutual information* between the input $X$ and output $Y$ pair for a channel, defined as

$$I(X:Y) = H(X) - H(X|Y) = H(Y) - H(Y|X), \tag{7}$$

where $H(X|Y)$ denotes the conditional entropy of $X$ given $Y$, defined as

$$H(X|Y) = -\sum_i p_i \sum_j q_{j,i} \ln q_{j,i}.$$

It can be shown that mutual information is symmetric and non-negative, $I(X:Y) = I(Y:X) \geq 0$. It represents the amount of information transmitted by the channel; this is reflected in the reduction of the uncertainty of the original source symbols H(X) via observing a related variable $Y$ (the channel output), quantified by $H(X|Y)$. The capacity of such a discrete, memoryless channel is given by maximizing $I(X:Y)$ over all possible input distributions $p_i$. In the context of PdPC, that amounts to the probability of $E$ being phosphorylated with, and without the kinase. An alternate interpretation of the mutual information (and hence channel capacity) is obtained by noting that $I(X:Y)$ is also a divergence measure between the *joint* probability mass function (p.m.f.) $p(x, y)$ of the input and output with the product of the input $p(x)$ and output $q(y)$ marginal p.m.f's, i.e.

$$I(X:Y) = \sum_{i,j} p(x_i, y_j) \ln \frac{p(x_i, y_j)}{p(x_i) q(y_j)} \tag{8}$$

This says that if the output $Y$ is independent of the input $X$, i.e., $p(x_i, y_j) = p(x_i) q(y_j)$, the output $Y$ provides no information about the input $X$, and hence $H(X|Y) = H(X)$. In all such cases, the channel carries no information, i.e. $I(X:Y) = 0$, and hence has zero capacity [16].

## 2.4 The capacity of a binary, non-symmetric channel

According to the information theory [13, 26, 5], a binary non-symmetric noisy channel is represented by the conditional probability matrix

$$\mathbf{K} = \begin{pmatrix} 1-a & a \\ b & 1-b \end{pmatrix}. \tag{9}$$

where $a = P\{Y = 1|X = 0\}$ and $b = \Pr\{Y = 0|X = 1\}$, $a \neq b$, represent the two errors introduced by the channel, and hence $a, b \ll 1$ is the regime of interest. Let us further assume the input signal has a probability distribution $(x, 1-x)$. Then the joint probabilities of input and output are the elements of

$$\begin{pmatrix} x(1-a) & xa \\ (1-x)b & (1-x)(1-b) \end{pmatrix}. \tag{10}$$

Using the divergence formulation in Eq. (8) above, the mutual information $I(X:Y)$ can be written in terms of the input p.m.f variable $x$ as follows:

$$\begin{aligned} I(x) &= x(1-a)\log_2 \frac{1-a}{x(1-a)+(1-x)b} + xa\log_2 \frac{a}{xa+(1-x)(1-b)} \\ &+ (1-x)b\log_2 \frac{b}{x(1-a)+(1-x)b} + (1-x)(1-b)\log_2 \frac{1-b}{xa+(1-x)(1-b)}. \end{aligned} \tag{11}$$

Hence, the channel capacity among all possible $x$ is obtained by

$$C = \max_{x \in [0,1]} I(x). \tag{12}$$

Some straightforward algebra leads to the optimal $x$

$$x^* = \frac{1-b(1+\sigma)}{(1-a-b)(1+\sigma)}, \tag{13}$$

in which

$$\sigma = \left( \frac{b^b(1-b)^{1-b}}{a^a(1-a)^{1-a}} \right)^{\frac{1}{1-a-b}}. \tag{14}$$

Therefore, the channel capacity is

$$C = I(x^*) = \log_2(1+\sigma) + \frac{1}{1-a-b}\log_2\left( \frac{a^{a(1-b)}(1-a)^{(1-a)(1-b)}}{b^{ba}(1-b)^{(1-b)a}} \right). \tag{15}$$

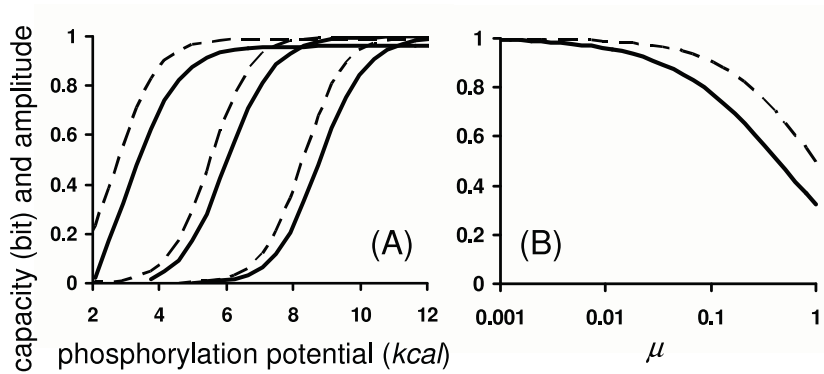

Figure 2: Channel capacity $C$ given in Eq. (15), and amplitude of switch ($AoS$) given in Eq. (4), of PdPC, as functions of $\Delta G = RT \ln \gamma$ and $\mu$. (A) Channel capacity (solid curves) and $AoS$ (dashed curves) from left to right with $\mu = 0.01, 10^{-4}, 10^{-6}$. (B) Channel capacity (solid curve) and $AoS$ (dashed curve) as function of $\mu$ with $\gamma = \infty$. One sees a general agreement between the channel capacity $C$ and $AoS$.

## 3 Information Theory of PdPC

### 3.1 Channel capacity $C$ and $AoS$ of PdPC

Applying the calculation for channel capacity, given in Eq. (15), to the information transfer matrix given in Eq. (5), we have

$$a = \frac{\mu}{1+\mu} \quad \text{and} \quad b = \frac{1}{1+\gamma\mu}. \tag{16}$$

Fig 2 shows the channel capacity $C$ as a function of $\gamma$ and $\mu$. In comparison, we also have drawn the amplitude of the switch ($AoS$) defined in Eq. (4) [21]. A general agreement between $C$ and $AoS$ is observed. Note that in terms of the $a$ and $b$ in Eq. (9), $AoS = 1 - a - b$. In fact, $AoS = 0$ or $a + b = 1$ if and only if $C = 0$ as can be readily verified from Eq. 11, when $\Delta G = 0$ (equivalently $\gamma = 1$).

Fig. 2A also shows that the mid-points of the dashed curves are all near their corresponding $\frac{1}{\mu}$. In fact, solving the mid-point of the $AoS$ from Eq. (4), we have the mid-point located at $\gamma_{0.5} = \frac{1+3\mu}{\mu(1-\mu)}$. So when the $\mu \ll 1$, $\gamma_{0.5} = \frac{1}{\mu}$. Therefore, the critical condition for the PdPC to function well as a communication channel is when $\gamma\mu \gg 1$. An insight from the above analysis is that, from the information transmission perspective, a PdPC is more energetically efficient, i.e., requires less amount of free energy expenditure, when $\mu$ is larger. The parameter $1/\mu$ is the equilibrium constant for the dephosphorylation reaction. For epidermal growth factor receptor (EGFR) the $\mu$ is indeed very large; it has been reported in the range of 0.5-1.6 [9].

Since, PdPC is a binary channel, the maximum information rate or capacity per channel use is $C = 1$ bit. This is in fact the implications in almost all the writing of cellular biologists: "When the kinase is activated, $E$ is phosphorylated; when the kinase is not activated, the $E$ is in its non-phosphorylated state." We now realize that this assumption implies that $\mu = 0$ and $\mu\gamma = \infty$.

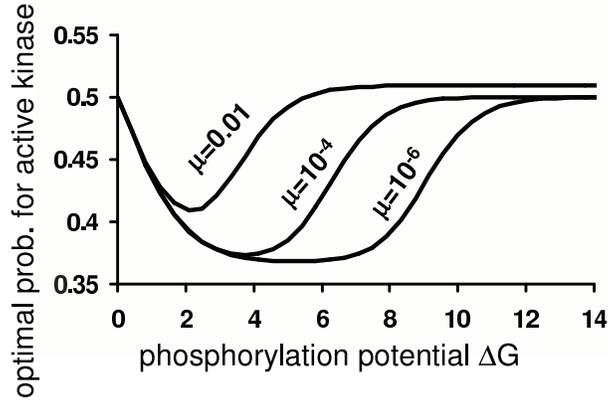

Figure 3: The optimal probability for kinase being active in order to fully utilize the channel capacity according to the information theory.

As we have stated, the channel capacity $C$ is defined for the optimal scenario of a channel usage. For the PdPC, the optimal usage is when the upstream kinase with probability $x^*$ being inactive, and $(1 - x^*)$ being active, where $x^*$ is given in Eq. (13). It is useful to determine the value of $x^*$ in the limit $\gamma \to \infty$ for a fixed $\mu$ as shown in Fig 3. Since $b = 1/(1 + \gamma\mu)$, it follows that $b \to 0$ when $\gamma \to \infty$ with fixed $\mu$. Furthermore from Eq. (14), we have

$$\lim_{\gamma \to \infty} \sigma = \frac{a^{-\frac{a}{1-a}}}{1-a} = \frac{(1+\mu)^{1+\mu}}{\mu^\mu}. \qquad (17)$$

Hence, in Eq. (13)

$$\lim_{\gamma \to \infty} x^* = \frac{(1+\mu)\mu^\mu}{\mu^\mu + (1+\mu)^{1+\mu}}. \qquad (18)$$

It follows that

$$\lim_{\mu \to 0} \lim_{\gamma \to \infty} x^* = \frac{1}{2}.$$

Fig. 3 shows that the optimal scenario for PdPC is when the kinase is little more than 50% of the time being active, and a little less than 50% of the time being inactive. In fact with $\gamma = \infty$, Eq. (18) shows that $1 - x^* = 0.509, 0.5002$, and $0.500003$ for $\mu = 0.01, 10^{-4}$, and $10^{-6}$ respectively.

## 3.2 Energy and information

Fig. 2A shows a general trend of increasing channel capacity with increasing free energy expenditure in the PdPC, from ATP hydrolysis. We also see that If $\Delta G = 0$, that is $\gamma = 1$ and $b = 1 - a$, then a changing of upstream kinase activity has no effect on the down steam protein phosphorylation. Hence $C = 0$. This result is well known in biochemistry: When a reaction is at its chemical equilibrium, changing the amount of enzyme, the kinase, can only change the rate process but not equilibrium concentrations [22, 2].

When the PdPC is kept away from chemical equilibrium, i.e. either with active ATP hydrolysis ($\gamma > 1$) or ATP synthesis ($\gamma < 1$), the mutual information between the upstream kinase and the downstream protein activities increases: The information then can be passed through the "channel". However, the greater mutual information transmission is achieved when $\gamma > 1$: This is indeed the regime the cell biology operates. One can further ask the amount of mutual information gain per unit of energy $\Delta G = RT \ln \gamma$:

$$\frac{dI}{d\Delta G} = \frac{\gamma}{RT}\frac{dI}{d\gamma} = -\frac{\mu\gamma(1-x)}{RT(1+\mu\gamma)^2}\log_2\left(\frac{b(xa+(1-x)b)}{(1-b)(x(1-a)+(1-x)b)}\right). \quad (19)$$

Fig. 4 shows the mutual information $I(x)$ (Eq. 11) increase with both increasing and decreasing from $\Delta G = 0$. It also shows that after $\Delta G > 5 kcal/mol$, the increase in the free energy leads to relatively little gain in $I$.

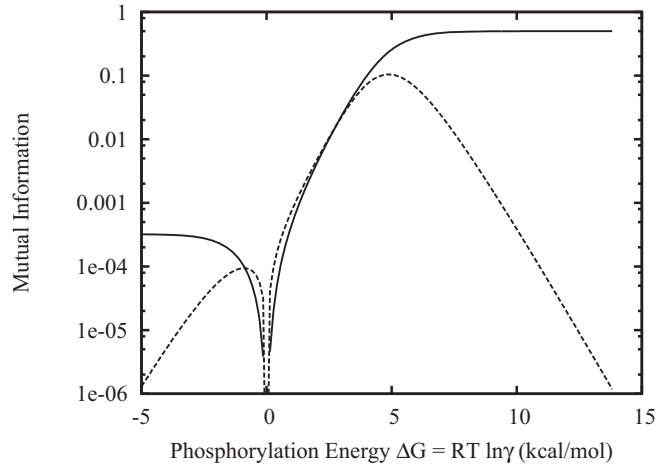

Figure 4: The mutual information $I$ between kinase activity and the substrate protein phosphorylation, shown by the solid curve. The dashed curve is the $dI/d\Delta G$, showing how much mutual information gain per unit of energy. The parameters for the computation: $\mu = 0.001$ and $x = 0.8$, i.e., the probability for upstream kinase being active is 0.2.

### 3.3 Sequential PdPC cascade and distributed code

We have discussed a single step PdPC in the previous sections. We now turn our attention to a sequence of PdPC. In current cell biology literature, this is widely known as "signaling cascade"

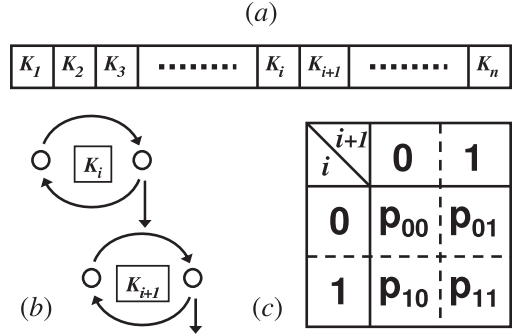

Figure 5: In cellular biochemistry, different PdPCs are often found to form a chain, called a signaling cascade pathway. This can be represented as (a). The relation between two successive kinases are shown in (b). The matrix in (c) quantifies the information transfer from a upstream kinase, $K_i$, to a downstream kinase $K_{i+1}$.

[24]. Fig. 5 illustrates a chain of the PdPC, each intermediate is a kinase itself which can be activated by its upstream kinase and activates its downstream kinase. For example, in the mitogen-activated protein kinase (MAPK) pathway, there are MAPKK(say Erk) and MAPKKK (say Raf) [28].

However, most discussions on signaling cascade emphasize the temporal aspect of the multiple PdPC: One mainly is interested in the relation between the first kinase activation event and the last protein phosphorylation in the chain. From information theoretical perspective, this means that one is interested in the information transfer matrix

$$\mathbf{K}^n = \begin{pmatrix} 1-a & a \\ b & 1-b \end{pmatrix}^n = \begin{pmatrix} 1 - \frac{a(1-\eta^n)}{1-\eta} & \frac{a(1-\eta^n)}{1-\eta} \\ \frac{b(1-\eta^n)}{1-\eta} & 1 - \frac{b(1-\eta^n)}{1-\eta} \end{pmatrix}, \qquad (20)$$

where $\eta = 1 - a - b$ is in fact the $AoS$ for single step PdPC. This is still a binary, asymmetric noisy channel. Note that since $\eta < 1$, the n-step $AoS$

$$AoS(n) = 1 - \frac{a(1-\eta^n)}{1-\eta} - \frac{b(1-\eta^n)}{1-\eta} = \eta^n \qquad (21)$$

decreases geometrically with $n$. This is reflected in Fig. 6, where the channel capacity decreases with the number of steps in the cascade and asymptotically approaches zero $\sim (a+b)^2 \eta^{2n}/(8ab)$. This is consistent with the well-known *data-processing inequality* in information theory [5] that states that any intermediate processing of data cannot increase mutual information.

With this information theoretic perspective, one naturally asks - why there are so many pathways with cascade? One widely given answer is that, with the multiple steps, a signaling process can have a multiple regulatory "entry point", and information is heavily integrated in the system through feedback. We certainly believe this is a very reasonable idea which we shall call "regulatability hypothesis". However, we would like to suggest an alternative perspective, the concept of distributed code [17]: If one moves away from focusing on the each and every kinase in the cascade chain one at a time, rather considers them together as a collection of different kinases that characterizes the states of the biochemical system, then it is not enough to know only the temporal sequential events of one activation after another. Rather, question such as "When $K_7$ is activated, is $K_3$ still activated, or no longer so?" becomes meaningful. Indeed, one should consider the $n$-dimensional binary variable $(K_1, K_2, \cdots, K_n)$ as the dynamic variable [6]. Such information is rarely provided in the experimental studies of cellular signal transduction.

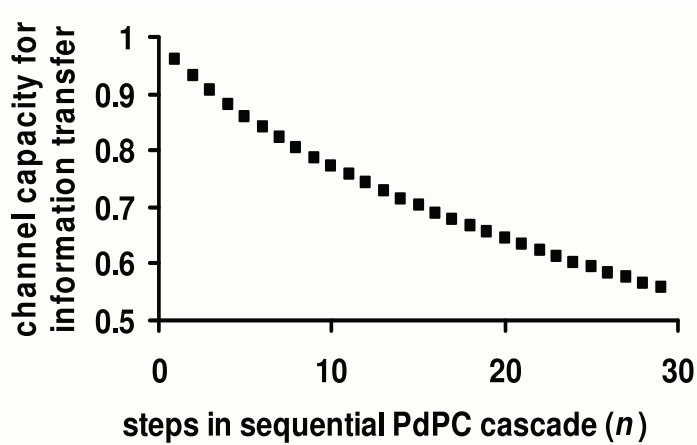

Figure 6: The channel capacity decreases with the number of steps in a sequential PdPC cascade, as shown in Fig. 5. Parameters used $\mu = 0.01$ and $\gamma = 5 \times 10^8$. The result illustrates the data-processing inequality in the information theory.

But this simultaneous information is crucial in understanding the function of a signaling cascade. For example, if one measures the fluctuating activity of a particular kinase as a function of time, such dynamics could contain deterministic oscillation or only stochastic fluctuations. One can not even addressed this question without at least simultaneously measuring a pair of relevant signaling molecules [27]. The biochemical state of a pathway is defined by the $n$-dimensional activation "pattern". We call this distributed code [17].

The distributed code and regulatability clearly are not exclusive. They are different functional aspects of a signaling cascade [19]. The former emphasizes the logical intra-relationship between the "players", while the latter emphasizes the system as a whole that define the biochemical states of a module. It is interesting to recognize that in information engineering, the source coding (or efficient information representation) and channel coding (or communication over a noisy channel), as reflected in Shannon's two major theorems, have been largely two separated communities. Signalling in biological cells suggests the need for a unification of both theories.

## 3.4 Markov Model for PdPC Encoding

One can quantify the possible amount of biochemical information encoded in such a distributed, multistage code (see Fig. 5a). Note that considering the distributed code is a very different question as the one treating the PdPC chain as multiple independent steps of a *memoryless* chain as in Eq. 20. Rather, the entire PdPC signaling cascade as a whole represents a code that follows a Markov chain whose one-step transition probability matrix is given by the stochastic matrix $\mathbf{K}$ in Eq. 5.

Because the correlation between the activations of a kinase and its substrate protein (i.e., the matrix in Eq. 5), not all the proteins can code completely independent information. Therefore, the information theory computes the entropy *per symbol* of an infinite long (Markov) sequence $\chi = \{X_n\}, n \geq 1$ is defined [5] as

$$H(\chi) = \lim_{n \to \infty} \frac{1}{n} H(X_1, X_2, \ldots X_n) \tag{22}$$

In biology, the infinitely long cascade does not make sense. Still, it is interesting, from biochemical information perspective, to obtain this parameter. It is know that for stationary, ergodic Markov processes, the entropy per symbol is given by

$$H(\chi) = \sum_i \pi_i \sum_j q_{j,i} \, ln q_{j,i} \tag{23}$$

where $q_{j,i} = P\{Y = y_j | X = x_i\} = \mathbf{K}_{i,j}$ are the 1-step transition probabilities of the chain given by the elements of the matrix $\mathbf{K}$, and $\pi_i$ are elements of the stationary (row) vector $\underline{\pi}$ of the chain that satisfies: $\underline{\pi} = \underline{\pi}\,\mathbf{K}$.

For the Markov transition matrix in Eq. (9), the stationary distribution $\underline{\pi}$ is readily computed as $[\pi_0, \pi_1] = \left[\frac{b}{a+b}, \frac{a}{a+b}\right]$. Hence the (asymptotic) Shannon entropy per symbol for the Markov chain, as a distributed multistage code, is:

$$\begin{aligned}
\widetilde{H} &= -\sum_{i=0,1} \pi_i \sum_{j=0,1} q_{j,i} \log_2 q_{j,i} \\
&= -\frac{(1-a)b\log_2(1-a) + ab\log_2 a + ab\log_2 b + a(1-b)\log_2(1-b)}{a+b} \\
&= \frac{(1+\mu)\log_2(1+\mu) + \mu\log_2 \gamma + \mu(1+\gamma\mu)\log_2\left(1+\frac{1}{\gamma\mu}\right)}{1+2\mu+\gamma\mu^2}.
\end{aligned} \tag{24}$$

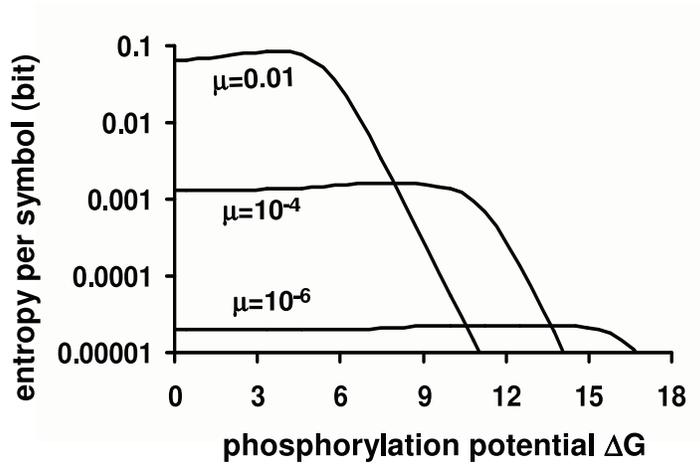

Figure 7: Entropy per symbol in a linear PdPC cascade as a multistage distributed code, $\widetilde{H}$, changing with $\mu$ and $\gamma$ according to Eq. (24).

Fig. 7 shows that as a multistage distributed code, a PdPC cascade has the entropy per symbol, $\widetilde{H}$, first increases with $\gamma$ and then eventually decreases. Note that to have a high entropy per symbol, one needs to have high entropy in the stationary distribution $\underline{\pi}$, as well as weak correlation between the successive stages. Fig. 7 shows that $\gamma$ increases the entropy of the $\underline{\pi}$ while also increases the correlation (i.e., mutual information). There is a competition between these two effects, leading to an optimal $\gamma\mu^2 \sim 1$.

For $\mu \ll 1$ and $\mu\gamma \gg 1$, we have $\widetilde{H} \approx \mu(2 + \log_2 \gamma)/(1 + \gamma\mu^2)$, which indeed first increases and then decreases with increasing $\gamma$.

# 4 Discussion

## 4.1 Information and free energy

In classical thermodynamics, energy is conserved. However, a spontaneous conversion of energy from one form to another, in real world, creats entropy. Entropy and free energy, in fact, were invented to account for the "usefulness" of energy. Ever since Shannon's work on entropy, there has been a continuous interest in the relation between information theory and thermodynamics, particularly in solid-state physics computers, in quantum theory, and in statistical mechanics [15, 3, 10].

Chemistry based cellular information processing provides another scenario where the fundamental relation, if any, between information theory and thermodynamics can and should be further investigated. In the present work, we are able to clearly show the following: *A communication channel consisting of a sequence of biochemical activities has zero capacity if and only if there is no free energy dissipation.* Furthermore, we have shown a clear positive correlation between the channel capacity and the amount of free energy dissipation (Fig. 2A). Whether these quantitative relations can be turned into something more universal remains to be further investigated.

## 4.2 Molecular structure, information storage, and biochemical communication

In molecular biology, since the discovery of DNA as the information storage for heredity, there has been a continuous fascination with Shannon's information theory [12, 25]. There are two theses in Shannon's information theory, which correspond precisely to his two theorems: The information storage, i.e., encoding, and the information transmission, i.e., channels. Our present analysis shows that in cellular signaling systems in terms of biochemical activities such as PdPC and GTPase, both issues are present: signaling cascade certainly delivers information as a relay channel, but it could also serves as a distributed multistage code, as recently suggested [19, 17]. As a thermodynamic system that processes information, we have shown that free energy expenditure is an indispensable part of biochemical communication inside a living cell.

## 4.3 Information, propensity and probability

Pursuing a possible unification of information theory and nonequilibrium thermodynamics must be based on the mathematical theory of probability with stochastic modeling. It is interesting to point out that three terms have been used in connection to the nature of uncertainties. The term information is often used by engineers. Philosophers, on the other hand, prefer to use the term "propensity" [7]. The most understood term however is probability, which are used by mathematicians. In some sense, the term information is from an observer's perspective [11], and the term propensity emphasizes the "intrinsic nature" of the observed. And the term probability has a more neutral connotation.

# 5 Acknowledgement

H.Q. acknowledges partial support from NSF grant No. EF0827592.